\documentclass{natureprintstyle} 

\usepackage{xcolor}
\usepackage{graphicx}
\usepackage{dcolumn}
\usepackage{bm}
\usepackage{color}
\usepackage{hyperref}
\usepackage{units}
\usepackage{hhline}
\usepackage{amsmath,amssymb}
\usepackage{float}
\usepackage{listings}
\usepackage[capitalize]{cleveref}
\usepackage{xspace}
\usepackage{subcaption}
\usepackage{xcolor}

\definecolor{codegreen}{rgb}{0,0.6,0}
\definecolor{codegray}{rgb}{0.5,0.5,0.5}
\definecolor{codepurple}{rgb}{0.58,0,0.82}
\definecolor{backcolour}{rgb}{0.95,0.95,0.92}

\lstdefinestyle{mystyle}{
    backgroundcolor=\color{backcolour},   
    commentstyle=\color{codegreen},
    keywordstyle=\color{magenta},
    numberstyle=\tiny\color{codegray},
    stringstyle=\color{codepurple},
    basicstyle=\ttfamily\footnotesize,
    breakatwhitespace=false,         
    breaklines=False,                 
    captionpos=b,                    
    keepspaces=true,                 
    numbersep=5pt,                  
    showspaces=false,                
    showstringspaces=false,
    showtabs=false,                  
    tabsize=2
}
\newcommand{\red}{\textcolor{black}}

\newcommand{\kookaburra}{\texttt{kookaburra}\xspace}
\newcommand{\bilby}{\texttt{bilby}\xspace}

\newcommand{\bpn}{B_{\rm p/n}}
\newcommand{\logbpn}{\log_{10}(B_{\rm p/n})}
\newcommand{\lnbpn}{\ln(B_{\rm p/n})}

\newcommand{\SPA}{School of Physics and Astronomy, Monash University, VIC 3800, Australia}
\newcommand{\OzGravMonash}{OzGrav: The ARC Centre of Excellence for Gravitational Wave Discovery, Clayton VIC 3800, Australia}

\usepackage{siunitx} 

\title{Flickering of the Vela pulsar during its 2016 glitch}

\author{Gregory Ashton$^{1, 2, 3}$, Paul D. Lasky$^{1, 2}$, Rowina Nathan$^{1, 2}$, and Jim Palfreyman$^{4}$}

\begin{document}
\maketitle

\begin{affiliations}
\item \SPA.
\item \OzGravMonash.
\item Royal Holloway, University of London, United Kingdom.
\item School of Natural Sciences,  University of Tasmania, Australia.
\end{affiliations}

\begin{abstract}
The first pulse-to-pulse observations of a neutron star glitch in the Vela pulsar identified a null pulse\cite{palfreyman18} hinting at the sudden disruption of the neutron star's magnetosphere. The only physical model connecting the glitch and the null pulse relies on a starquake either triggering, or being triggered by, the glitch itself\cite{Bransgrove2020}.  Until now, this was the only null pulse identified from over 50 years of observing the Vela pulsar\cite{Johnston2001,palfreyman18}. We identify five other null-like pulses, that we term \textit{quasi-nulls}, before and after the glitch, separated by hundreds of seconds. We verify that such nulls are not found in data away from the glitch. We speculate that the quasi-nulls are associated with foreshocks and aftershocks preceding and following the main quake, analogously with terrestrial quakes. This implies the energy reservoir built up between glitches is not released suddenly, but over a period of minutes to hours around the time of the glitch.  
\end{abstract}

In 2016, the Vela radio pulsar increased its rotation frequency by one part in a million\cite{palfreyman18}. Such events, known as glitches, provide rare glimpses into the internal structure of neutron stars. 
Perhaps most intriguingly, a series of unexpected changes in the pulsations were observed coincident with the Vela glitch.
First, a broader-than-usual pulse, pulse 76\footnote{We follow the notation and numbering of the original analysis\cite{palfreyman18} in which pulse $n$ refers to the $n$th set of period-folded data. This naming means that if in the $n$th data set the pulsar nulls (i.e. no pulsation is observed), it is still referred to as pulse $n$.}. Second, during pulse 77, the emission disappeared, a phenomenon known as a null\cite{Backer1970}. These two anomalous pulses were then followed by two pulses with lower-than-usual linear polarization and several pulsations which appeared to arrive slightly later than expected.
Individual pulses from radio pulsars typically show a high degree of variability and the Vela pulsar is no exception, but before pulse~77, it has never been known to have nulled\cite{Biggs1992, Johnston2001}.

Understanding the behaviour of the Vela pulsar during its glitch has significant scientific value. The glitch itself is likely caused by a sudden coupling of the superfluid interior to the crust of the star\cite{haskell2015}. At the same time, the abrupt changes in the pulses likely arise from changes in the magnetosphere. The Vela glitch offers the tantalizing opportunity to simultaneously understand more about both the interior of the neutron star and the magnetosphere.
The broadening and null can be explained by a ``quake quenching'' episode\cite{Bransgrove2020}: a crust quake, which disrupts the magnetosphere causing the broadening and null, then triggers the superfluid unpinning leading to the observed spin up.

\red{
In addition to the unusual broadening and null, we have previously found evidence\cite{ashton2019} that the glitch is not a simple spin-up event. We find the rotation frequency of the star ``overshoots'', suggesting the existence of three distinct components to the neutron star\cite{graber18}. Modelled fits to the data\cite{pizzochero2020,montoli2020} confirm that a three-component model is able to explain the observed overshoot and infer the physical parameters.
We also discovered the hitherto unseen phenomenon of a slow down in the rotation rate prior to the glitch\cite{ashton2019} (also consistent with a magnetospheric slip\cite{montoli2020}). While the evidence for the overshoot is overwhelming, the slow-down before the glitch is tentative. Piecing this rich behaviour together into a unified model holds great promise for delivering new insights into neutron star physics. Ref\cite{erbil2020} propose that the cooperative action of crust breaking and vortex pinning-unpinning may be responsible, yielding predictions for the next glitch time to within a day. Future observations are clearly needed to decide if these predictions are born out.
}

In this work, we perform a systematic study of the individual pulses around the time of the 2016 glitch. We set out to answer two questions: 1) What is the statistical significance of the broadening and null pulse identified in Ref.\cite{palfreyman18}? 2) Are there pulse-shape changes during the glitch which could help us understand the observed pre-glitch slow-down? In the following sections, we give an overview of our method, address each of these questions in turn, then provide a discussion and outlook.

\section*{Shapelet-based pulse characterisation}

The initial discovery\cite{palfreyman18} of the magnetospheric alterations during the 2016 Vela pulsar glitch analysed the data using a modification to traditional pulsar timing methods\cite{palfreyman:thesis}. A high signal-to-noise ratio (SNR) template is matched-filtered against individual pulses to identify the pulse time of arrivals (TOA), defined at the maximum flux of the template. (This approach differs from standard usage where interest is in the long-term timing of a pulsar, in which case the matched-filter is applied to an averaged pulse\cite{Lyne1988}).

In this work, we fit a shapelet pulse-model\cite{Refregier2003, Lentati2015} to $\sim33,000$ individual pulses around the glitch. 
We provide an overview of the \texttt{kookaburra} package used in the Methods section.
To compare to the regular behaviour of the pulsar, we also analyse $\sim13,000$ pulses in a data set taken 85 days before the glitch.
We pre-process the raw data using the \texttt{PSRCHIVE} package\cite{psrchive}, performing a frequency and polarisation ``scrunch'' and removing the baseline flux.
For each pulse, we fit the shapelet flux model with a maximum of six components along with a polynomial model of degree two for the base flux. This enables us to model any remaining base-flux fluctuations not removed in pre-processing. The shapelet flux model is parameterised by the pulse width $\beta$, the shapelet coefficients $C_{i}$ (where $i \in [0, 5]$), and the pulse time of arrival $t_{i}$.
Our analysis yields posterior distributions for these shapelet and base-flux parameters. We additionally analyse each pulse using a null model consisting of the base-flux model only. Comparing the evidence (the fully marginalized likelihood) for the pulse and null model yields the Bayes factor $\bpn$ comparing the pulse and null models. In the Methods section, we provide a discussion of the Bayes factor and the sensitivity of the method to finding nulls.

\section*{Statistical significance: Identifying quasi-nulls}

\begin{table}[tb]
    \centering
    \begin{tabular}{l|l|l|l|}
         Pulse & $\log_{10} B_{\rm p/n}$ & $\beta$ [ms] & SAT [s] \\ \hline
         -15433$^*$ & -0.2 & $1.2\pm 0.5$ & $-1388.13\pm0.001$ \\
         -1167$^*$ & -1.5 & $2.7\pm 0.9$ & $-111.378 \pm 0.0009$\\
         9$^*$ & -0.5 & $0.7 \pm 0.5$ & $-6.07792\pm0.0008$ \\
         76 & 117 & $5.2 \pm 0.5 $ & $-0.0885894 \pm 0.002$\\
         77 & -10.9 & -- & 0 \\
         4609$^*$ & 0.2 & $0.5\pm 0.4$ & $405.74\pm0.002$\\
         9526$^*$ & 0.8 & $1.0\pm0.4$ & $845.804\pm0.001$ \\
    \end{tabular}
    \caption{The pulse-to-null Bayes factor $\logbpn$, pulse width $\beta$, and site arrival time (relative to the null pulse) for the outlier pulses observed in \cref{fig:onsource_bayes_factor}. The pulse width is not measured for the null pulse, pulse 77. Quasi-nulls are marked with an asterisk.}
    \label{tab:pulses}
\end{table}

\begin{figure*}[t]
    \centering
    \begin{subfigure}[b]{0.49\textwidth}
    \centering
    \includegraphics{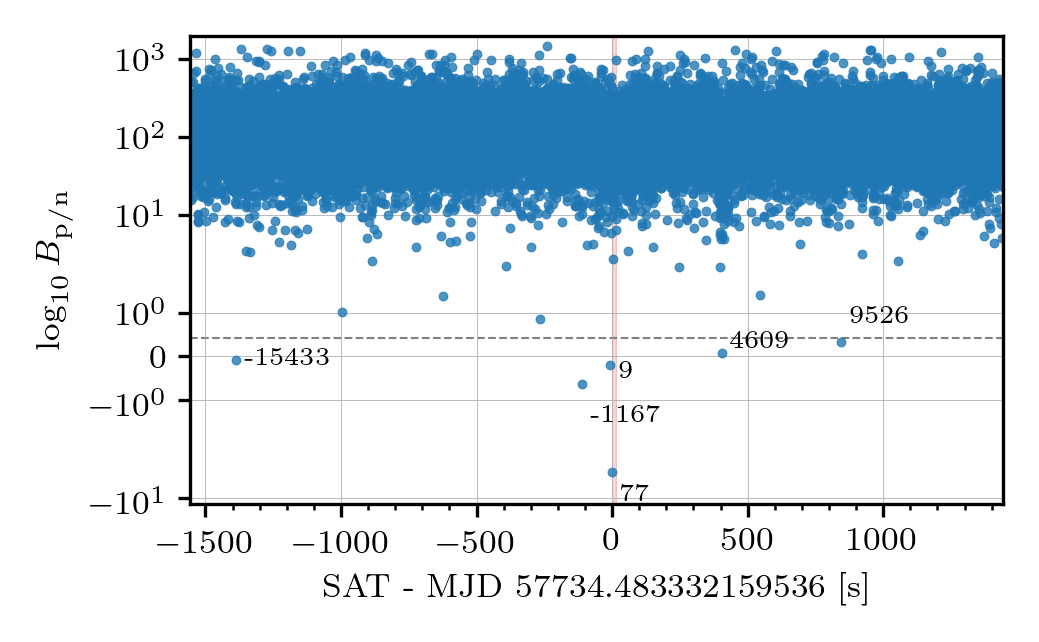}
    \caption{Data during the glitch; all times are referenced to pulse 77, the null pulse\cite{palfreyman18}.}
    \label{fig:onsource_bayes_factor}
    \end{subfigure}
    \hfill
    \begin{subfigure}[b]{0.49\textwidth}
    \centering
    \includegraphics{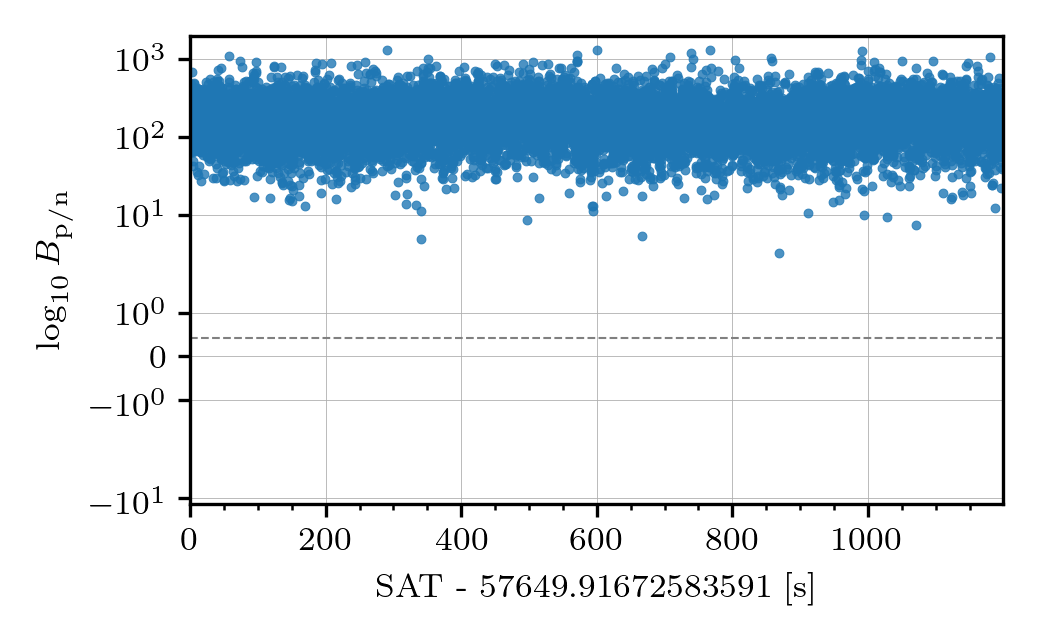}
    \caption{Data from MJD~57649, 85 days before the glitch.}
    \label{fig:offsource_bayes_factor}
    \end{subfigure}
    \caption{The pulse-to-null Bayes factor $\bpn$ against the median inferred site arrival times (SAT) at the Mount Pleasant \unit[26]{m} observatory. If $\logbpn{\gtrsim}1$ (as is the case for typical pulses), the represents strong evidence for a pulse rather than a null; if $\logbpn {\lesssim} -1$ this represents strong evidence for a null rather than a pulse; intermediate values indicate a weak preference either way. A shaded red region in Figure (a) indicates the $\sim 13$~s interval of the inferred\cite{ashton2019} glitch-time. \red{A horizontal dashed lines indicates our $\lnbpn < 1$ quasi-null threshold.}}
\end{figure*}

Understanding the significance of the unusual behaviour of the magnetosphere is paramount to drawing physical conclusions\cite{Bransgrove2020}. We can probe the significance of the null by comparing the Bayes factor $\bpn$ of pulse 77 with that of the other pulses in the data set.

In \cref{fig:onsource_bayes_factor}, we plot the pulse-to-null Bayes factor for the data surrounding the glitch, \cref{fig:onsource_bayes_factor}, and the data set 85 days before the glitch, \cref{fig:offsource_bayes_factor}. For almost all pulses, $\logbpn > 1$, i.e. strong evidence for a pulse rather than a null. Pulse 77 is the most significant null in the data set, reaffirming the initial findings\cite{palfreyman18}. 
Pulse 76 is not a significant outlier in terms of the pulse-to-null Bayes factor. However, it is clearly distinguished in our analysis as an outlier in the measured pulse width $\beta$ (see \cref{tab:pulses}).

\begin{figure*}
    \centering
    \begin{subfigure}{.5\textwidth}
    \includegraphics{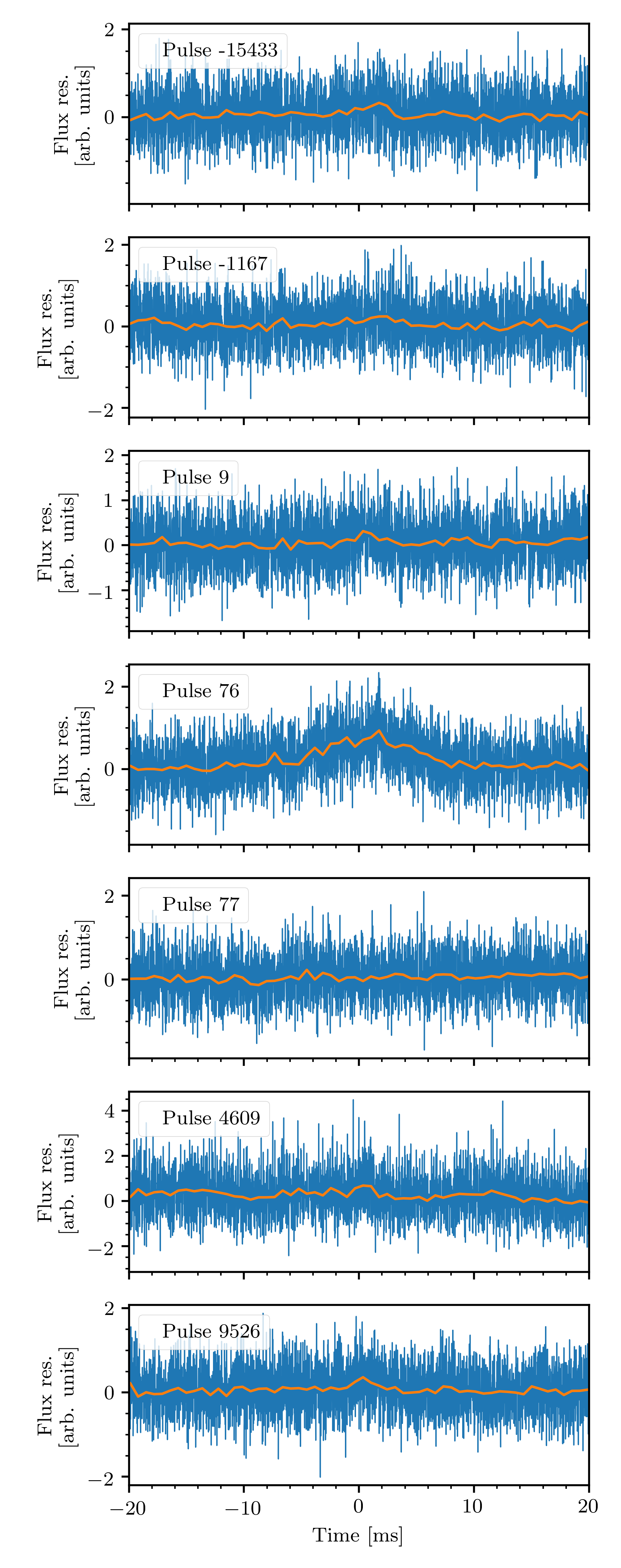}
    \caption{}
    \label{fig:pulses_a}
    \end{subfigure}%
    \begin{subfigure}{.5\textwidth}
    \includegraphics{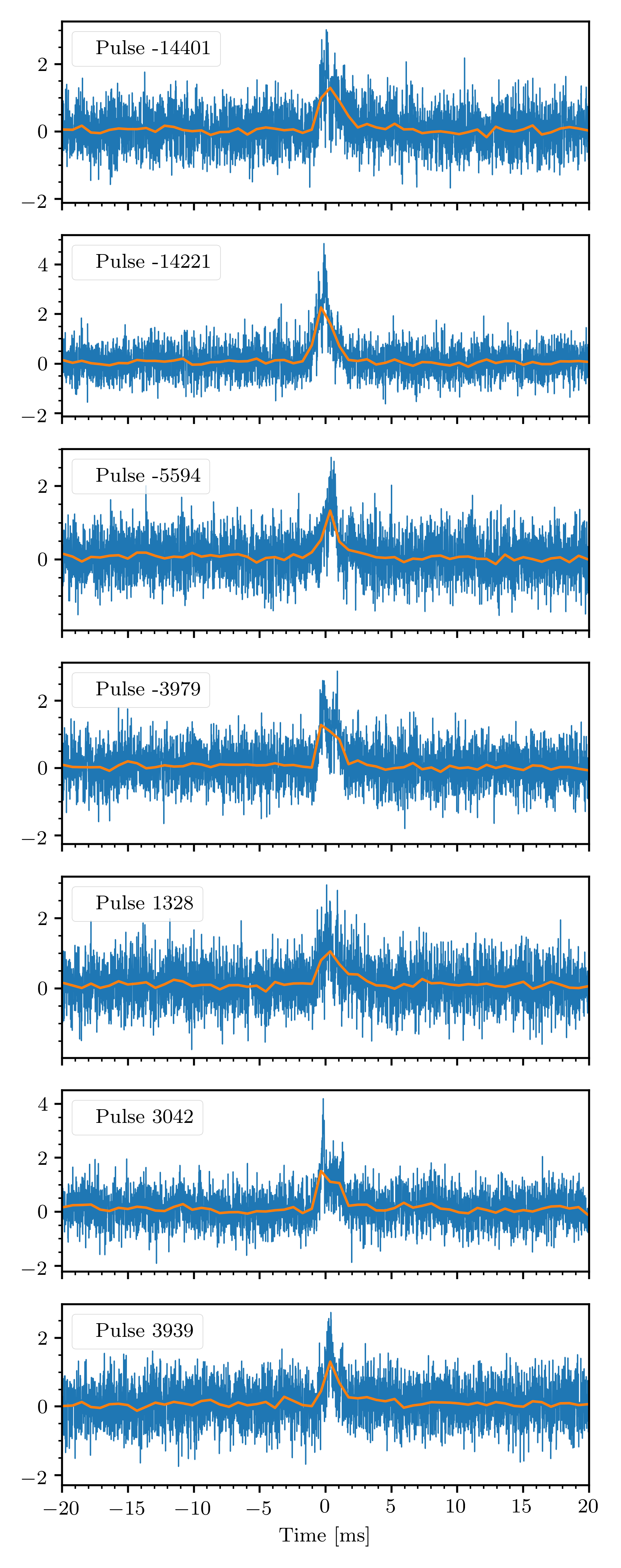}
    \caption{}
    \label{fig:pulses_b}
    \end{subfigure}%
    \caption{(a) The flux residual after removing base-line flux pulse data (blue) and running average (orange) for the outlier pulses identified in \cref{tab:pulses}. (b) Flux residuals for seven randomly selected pulses, used to place the outliers in context.}
    \label{fig:pulses}
\end{figure*}

Surprisingly, in addition to pulse 77, several other pulses can be identified in \cref{fig:onsource_bayes_factor} with $\lnbpn < 1$. We tabulate the pulse numbers and properties of these in \cref{tab:pulses} and, in \cref{fig:pulses_a}, we visualise the raw data. (We also include pulse 76, the broad pulse, in the set for completeness). We refer to the anomalous pulses with $\lnbpn < 1$ as \emph{quasi-nulls}: pulses which show some small pulse (\cref{fig:pulses_a}), but have a Bayes factor distinct from the bulk of the population (\cref{fig:onsource_bayes_factor}). \red{The quasi-null threshold of $\lnbpn < 1$ is arbitrary and we note that there are several other pulses which sit clearly in tails of the bulk of the distribution. We choose a conservative threshold to identify the subset of quasi-nulls in \cref{tab:pulses} for further analysis.}

To put the typical properties of the quasi-nulls into context, in \cref{fig:pulses_b} we provide the raw data for an equal number of randomly-selected pulses from the data surrounding the glitch. This clearly demonstrates the distinct nature of the anomalous pulses (which include the broad pulse 76 and nulling pulse 77) from that of randomly selected pulses. The additional anomalous pulses (pulse -15433, -1167, 9, 4609, and 9526) do not show strong evidence for or against a pulsation. This is reflected in that the Bayes factor close to zero. However, compared to typical pulses, they are clearly much weaker.

Are quasi-nulls a regular occurrence in the Vela pulsar? To determine if this is the case, we randomly select a set of data 85 days before the glitch. We repeat our analysis and, in \cref{fig:offsource_bayes_factor}, present the pulse-to-null $\bpn$ for each pulse.
Unlike the data surrounding the glitch, we observe no Bayes factors with $\lnbpn < 1$, i.e. no quasi-nulls. Taking an approximate rate of one per 200 seconds for the quasi-nulls in \cref{fig:bayes_factor} \red{(based on the quasi-nulls identified in \cref{tab:pulses} and the other outliers in \cref{fig:onsource_bayes_factor})} and assuming quasi-nulls arise from a Poisson process, the probability of seeing zero quasi-nulls in the $\sim1200$~s of off-glitch data is 0.0024 (a $3\sigma$ Gaussian-equivalent).

Why does our shapelet-based method identify these quasi-nulls when they were not identified in the original analysis\cite{palfreyman18}? The original analysis
identifies a null by a failure to calculate the TOA. That is, if a matched-filter analysis attempting to identify the TOA fails, the pulse is flagged as a potential null and flagged for follow-up\cite{palfreyman:thesis}. The quasi-nulls do have some pulse structure (see \cref{fig:pulses_a}) and therefore will not be flagged by this failure-to-identify-a-TOA method. By comparison, the shapelet-based method used here effectively quantifies the SNR of the pulse; thus giving a scale of how loud the signal is rather than a simple binary choice between a pulse and a null. This difference has allowed us to identify the quasi-nulls in the data set.

\section*{Systematic pulse-shape changes}

The second question this work sets out to answer is if there is evidence for systematic pulse-shape changes during the glitch. The initial analyses\cite{palfreyman18, ashton2019} used the traditional method of estimating the pulse TOAs: matched filtering a high-SNR template against the data. If the pulses contain systematic changes in their shape (either in a slowly-varying manner, or as sudden sharp features), these will manifest as changes in the timing of the pulsar. As such, the pre-glitch slow-down\cite{ashton2019} could actually be due to systematic changes in the pulse shape. Our systematic study of the pulse shape allows us to study this.

Visualisations of the median values of pulse-shape parameters (the pulse width and shapelet coefficients) do not yield any evidence of systematic changes in the pulse-shape during the glitch.

Comparing the time of arrivals measured using the shapelet-based method with that of the traditional timing method\cite{palfreyman18}, we also do not find any evidence of systematic changes in the inferred time of arrivals. Specifically, the median values from the two methods agree to within the stated uncertainties. We note that the timing uncertainty of the shapelet-based method is much larger than that of the traditional matched-filtering method. This can be understood as a statement that the arrival times measured by the shapelet-based method marginalizes over the uncertainty in the pulse shape, while the traditional method has only the timing uncertainty in the fit of the high-SNR template.

\section*{Discussion and Outlook}
During the 2016 glitch, the Vela pulsar appears to flicker: we identify five quasi-nulls---pulses with significantly reduced amplitude compared to regular pulses. We do not find any quasi-nulls in data taken 85 days before the glitch. This implies that quasi-nulls, like the null and pulse broadening\cite{palfreyman18}, may be related to the glitch itself.

We analyse $\sim33,000$ pulses ($\sim 2900~s$ of data) and can estimate the recurrence rate of quasi-nulls to be on the order of hundreds of seconds. However, the duration of data studied (relative to the recurrence rate) is insufficient to determine the waiting-time distribution (e.g., is it a Poisson process or periodic?) or constrain the total duration of time over which the pulsar is afflicted by quasi-nulls. To answer these questions, we need to analyse a much larger data set ($\gtrsim 10^{4}$~s) which, with our current shapelet-based method (and stochastic sampling) is computationally demanding. In future work, we will develop a matched-filter approach  to identify quasi-nulls that can be be applied to the required duration of data. This will enable us to study the behaviour of the pulsar in the days and weeks leading up to and after the glitch and confidently determine when the quasi-nulls start and stop.

If we can confidently associate quasi-nulls with the glitch itself, this may mean that a glitch could be predicted ahead of time by real-time pulse monitoring. An optimised matched filter system, consuming the data as it is recorded, could provide an early-warning alert that a glitch is imminent. This will enable a slew of instruments to observe the Vela pulsar and potentially capture the glitch in significantly more detail than is currently possible.

In addition to finding quasi-nulls, we also reaffirm the existence of the broad pulse 76 and the null pulse 77. We find no evidence of systematic changes in the pulse shape which could explain the pre-glitch slow-down\cite{ashton2019}. 

Our findings have deep implications. The quake-quenching mechanism\cite{Bransgrove2020} posits that the existence of the one magnetospheric anomaly (i.e., broad pulse 76 and null pulse 77) at the time of the glitch necessitates a crust-cracking episode to causally link the two phenomena. In other words, if the glitch occurs because of a mechanism internal to the star such as vortex unpinning, then the only way to simultaneously trigger a change in the magnetosphere geometry as inferred by the broad and null pulses is through a starquake. So then what of these quasi-null pulses before and after the glitch? 

It is reasonable to assume the quasi-nulls are caused by changes in the geometry of the star's external magnetic field, in the same way as hypothesized\cite{palfreyman18} for pulses 76 and 77.
But what causes these changes in the magnetosphere? One can speculate that such changes could also be caused by a crustal quake, implying the pre- and post-glitch quasi-nulls are respectively foreshocks and aftershocks akin to terrestrial tremors associated with large earthquakes. In this sense, there is a build up of stress inside the core or crust of the star during the interglitch period (i.e., the approximate three-year period between Vela glitches); when the stress is released, it happens sporadically over a series of events, with the main starquake also being associated with the glitch. In principle, understanding the statistics of the waiting-time distribution for the quasi-nulls could provide insight into this speculation.

If foreshocks and aftershocks explain the quasi-nulls, it is difficult to imagine that the stress builds up in the core of the star. One would require mini-vortex avalanches that carry enough stress to crack the stellar crust, but not enough energy to change the rotational evolution of the star, at least at the same level as the glitch. In that case, the stress must be being built up in the crust itself. This has an interesting consequence. If stress is being built up in the crust of the star, then it must be the crust quake that \textit{causes} the glitch itself, rather than an internal stellar mechanism (such as large-scale vortex unpinning) that causes the crust quake. In this way, it would be the crust that breaks, triggering the vortex avalanche that causes the glitch itself.  

Perhaps the quasi-nulls are \textit{not} associated with crust quakes at all. One can imagine that Vela is in a heightened state of magnetic activity for a period of time around the glitch.
This would imply that the order of causality is reversed from the quake-quenching scenario: the magnetic activity is required to cause a break in the crust, and thereby cause the glitch itself. While this interesting, we find it difficult to imagine how such can explain the quasi-periodic timing of Vela glitches. While we do not prefer this model at this time, we also cannot rule it out.
In either of the discussed scenarios, one would need to explain two disparate timescales: the short timescale of each null and quasi-nulls, and the much longer timescale between any two successive nulls.

\red{
We propose two speculative models, both of which require modeling to understand whether they are plausible, and further data analysis to understand if they are compatible with the observations. There are several observational prospects for furthering our insight into the Vela glitch. First, extended analysis of the unique pulse-to-pulse data surrounding the 2016 glitch, which can potentially constrain the duration and waiting-time distribution of the quasi-nulls. Second, analysis of pulse-to-pulse off-glitch times (both from the Mount Pleasant Radio Observatory and other, more sensitive, instruments) will determine decisively if quasi-nulls are unique to glitch epochs. Third, analysis of previous glitches\cite{dodson2002, dodson07}; while individual pulses are not resolvable in this data, we may nevertheless be able to apply statistical methods to determine if these glitches are accompanied by quasi-nulls. Finally, there are tentative predictions for gravitational waves accompanying the glitch\cite{yim2020} which, if detectable by ground-based gravitational-wave detectors, would provide a unique new view of the glitch.  In summary, further modeling and analysis of data are required, but offer exciting prospects for insights into the fascinating phenomenology and physics behind the quasi-nulls identified herein.}

\section*{Data availability}
The raw data used in this work is available from Ref.\cite{palfreyman18}. The results presented in this work (summary statistic and Bayes factors for each pulse) are available from Ref.\cite{zenodo_results} and can be read in using the \texttt{pandas} python package\cite{pandas:2010, pandas:2020}:
\begin{lstlisting}[language=python]
>>> import pandas as pd
>>> df = pd.read_hdf("kb_database.h5")
\end{lstlisting}

\section*{Code availability}
The results presented in this work are generated using the \texttt{kookaburra} package \url{https://kookaburra.readthedocs.io/}. The specific command-line used to analyse pulses in this work is
\begin{lstlisting}[language=bash, caption={}, label=listing:code]
kb_single_pulse all_data.h5 -p ${PULSE_NUMBER} \
    -s 6 -b 2 \
    --c-mix 0.1 \
    --beta-max 1e-7 --beta-min 1e-9 --beta-type log-uniform \
    --c-mix 0.1 --truncate-data 0.2 \
    --toa-prior-time 0.5 --toa-prior-width 0.1 \
    --sampler pymultinest --nlive 2000 \
\end{lstlisting}
Here the flux consists of a single shapelet with six components and a polynomial base-flux model of degree two; the shapelet coefficient is a slab-spike prior with mixture-ratio $\xi=0.1$ (the default in \kookaburra, see \cref{eqn:slab-spike} for the definition); the $\beta$ pulse-width parameter (see \cref{eqn:shapelet}) is given a log-uniform prior over the range $10^{-7}$ to $10^{-9}$ days;  the data is truncated to 20\% of the full rotation period (discarding data which does not contain the pulse itself); the prior on the arrival time is centered and contains 10\% of the full rotation period; and we use the \texttt{pymultinest} sampler\cite{buchner:2014}.

\section*{References}
\bibliography{bibliography}

\section*{Acknowledgements}
The authors thank Ashley Bransgrove and Yuri Levin for useful discussions and comments during the preparation of this work. We also thank Alessandro Montoli, 	Marco Antonelli, and Garvin Yim, for useful feedback improving the presentation of the work.
All computing was performed on the OzSTAR Australian national facility at Swinburne University of Technology, which receives funding in part from the Astronomy National Collaborative Research Infrastructure Strategy (NCRIS) allocation provided by the Australian Government,
PDL is supported through Australian Research Council Future Fellowship FT160100112, ARC Discovery Project DP180103155, and ARC Centre of Excellence CE170100004.
\kookaburra builds on an extensive software stack and the authors appreciate the
efforts of the community to build a vibrant ecosystem which makes work such as
this possible. \kookaburra uses \texttt{scipy}\cite{scipy:2020}, \texttt{numpy}\cite{oliphant:2006},
and \texttt{pandas}\cite{pandas:2010, pandas:2020} for data handling and manipulation;
\texttt{matplotlib}\cite{hunter:2007} for visualisation; \texttt{bilby}\cite{bilby} for
Bayesian inference-related aspects. All results in this work were generated
using the \texttt{pymultinest} sampler\cite{buchner:2014}.

\section*{Contributions to the paper}
G.A. was responsible for the data curation and analysis. G.A., P.D.L., R.N, and J.P were responsible for the interpretation and discussion. J.P. was responsible for the initial data collection and reduction.

\section*{Competing Interests}
The authors declare no competing interests.

\section*{Additional information}
\textbf{Correspondence and requests for materials} should we addressed to G.A, \href{gregory.ashton@ligo.org}{gregory.ashton@ligo.org}.

\begin{methods}
This work makes use of the python package \texttt{Kookaburra} which provides methods to fit flux models to individual pulsations. Fitting is performed by stochastic sampling methods using the \texttt{bilby}\cite{bilby} Bayesian inference library. \texttt{Kookaburra} offers both a command-line executable and highly flexible python Application Programming Interface (API). In this Methods section, we detail the implementation and verification of the \texttt{Kookaburra} software.

\textbf{The flux model:}
The primary flux model provided by \kookaburra is a simplified version of the
shapelet model\cite{refregier:2003}. We define a shapelet flux model
\begin{align}
    f(t) = \sum_{i=0}^{n_s} C_{i} H_{i}(t/\beta) e^{-t^2 / \beta^2} \,,
\label{eqn:shapelet}
\end{align}
where $C_{i}$ are the shapelet coefficients, $H_{i}$ is the Hermite polynomial
of degree $i$, and $\beta$ is a width parameter. Our modification simplifies
the definition of the coefficients, removing, in particular, a pre-factor of
$1/\beta$ and resulting in an orthogonal, but not orthonormal basis. The
complete single-component shapelet flux model fit to the data is then $f(t -
\tau)$ where $\tau$ is the pulse time of arrival. \kookaburra provides the
option to fit multiple additive shapelet flux models at the same time with a
prior on the time of arrival uniformly distributed between the components.

In addition to the pulse itself, radio-pulsar data usually contains a background flux.
This background flux can be removed using the \texttt{PSRCHIVE} package\cite{psrchive} (so-called baseline removal). In \kookaburra, we provide an alternative: we model the background flux by a polynomial ``base-flux model'' of arbitrary
degree with reference time centred to the middle of the observation. Adding this to the shapelet-flux model, the
complete flux model used in fitting is therefore
\begin{align}
F(t) = \sum_{i=0}^{n_p} B_{i}(t - t_{\rm mid})^{i} + \sum_{j=0}^{n_m} f(t - \tau)\,.
\label{eqn:full_flux}
\end{align}
where $n_p$ is the degree of the base polynomial and $n_m$ is the number of
shapelet models (each having an independent number of shapelet components
$n_s$).

Users who preprocess data with the \texttt{PSRCHIVE} baseline removal can turn off the \kookaburra base-flux model or they may wish to use it to capture any residual base-flux not removed in preprocessing (as done in the main body of this work). In our experience, it is preferable to use at least a linear polynomial base-flux model; this ensures any residual base-flux will not bias the inferred shapelet model.

The flux model described in \cref{eqn:full_flux} is implemented in the executable \texttt{kb\_single\_pulse}.
Generalisations can be made by using the underlying python API and extending the set of flux models in ``kookaburra.flux''.

Once the flux model is defined, a stochastic sampling algorithm (accessed via
\bilby\cite{bilby}) is used to fit the model to the data assuming a Gaussian likelihood (i.e. the flux is modelled as a sum of the deterministic model and a random Gaussian noise process). Exact details of the likelihood and extensions can be made in the \texttt{kookaburra.likelihood} module.

\textbf{Pulse and Null models:}
In analysing a set of data, we can estimate the probability of a null by
running an analysis excluding the components of the flux model intended to
model the pulsation itself. Using the polynomial base flux, our null model is

\begin{equation}\label{eqn:null_flux}
F(t) = \sum_{i=0}^{n_p} B_{i}(t - t_{\rm mid})^{i}\,.
\end{equation}

Using stochastic sampling, we can fit \cref{eqn:full_flux} and \cref{eqn:null_flux}, the difference in
log-evidences obtained from each constitutes a Bayes factor $B_{\rm p/n}$
quantifying the probability the data contains a pulsation vs. a null. For typical individual pulses, the shapelet model captures the rich features of the pulse leading to $B_{\rm p/n} \ll 1$. Indeed, for folded pulsations, the Bayes factors can become sufficiently large that nested-sampling based approaches can be slow to produce a solution. We are developing and testing alternative methods which will reduce the time-to-solution for these cases. For data not containing a pulse (i.e. off-pulse or null pulses), the shapelet model will typically capture some arbitrary feature of the noise. In these cases, the improvement in fit afforded by the shapelet flux is small and overwhelmed by the larger prior odds for the shapelet and base-flux model (this is also known as the Occam factor\cite{mackay:2003}). The result is a Bayes factor favouring the null pulse, $B_{\rm p/n} < 1$.

\textbf{Slab and Spike priors}
The flux-model coefficients, $C_i$ determine the contribution of each term in
the shapelet model to the overall flux. \kookaburra uses a so-called slab-spike
prior (see, e.g. Ref.\cite{malsiner:2018}). This a mixture-model prior given by

\begin{align}
    \pi(C_i) = \left\{ \begin{array}{cc} \xi & \textrm{ if } C_i = \hat{C}_i \\ (1-\xi)\pi'(C_i)  & \textrm{otherwise} \end{array}\right.\,,
    \label{eqn:slab-spike}
\end{align}
where $\xi$ determines the mixing fraction between the spike $\hat{C}_i$ and
the slab $\pi'(C_i)$. We then choose the spike to be at zero (i.e. the
contribution from the $i^{\rm th}$ component is zero) and the slab to be a
uniform distribution from zero to a maximum value.
We find that the slab-spike prior greatly improves the performance of
the stochastic sampling algorithm as it can ``turn off'' components which do not
improve the fit to the data.

\textbf{Maximum number of components:}
The optimal (in the sense of maximising the Bayesian model evidence) number of
shapelet models $n_m$, the number of components for each model $n_s$, and the
degree of the base polynomial $b_p$ is unknown for any data set (except in the
case of simulated data). This means we have uncertainty about the model-space
dimensionality. Typical stochastic sampling algorithms require a
fixed-dimensional space to satisfy their underlying assumptions and ensure the
results are a proper reflection of the posterior distribution and evidence.
Reversible-jump MCMC (RJMCMC) methods\cite{green:1995} enable trans-dimensional
sampling and a posterior estimate of the dimensionality and the components of
each dimension, marginalized over the full uncertainty. However, RJMCMC methods
typically require specialised implementations for the problem in hand.  When
faced with an unknown model dimensionality, a cheap and effective alternative
to implementing an RJMCMC sampler is to run identical analyses, but varying the
model dimensionality. For \kookaburra, we find that setting a sufficiently large
number of components (determined experimentally) combined with the slab-spike
priors result in an efficient sampling of high-dimensional spaces.  The
maximum number of components will depend on the analysis at hand, in the
literature, values as large as 30 are typical\cite{lentati:2017} for radio pulsar
profiles. 

\begin{figure}
    \centering
    \includegraphics{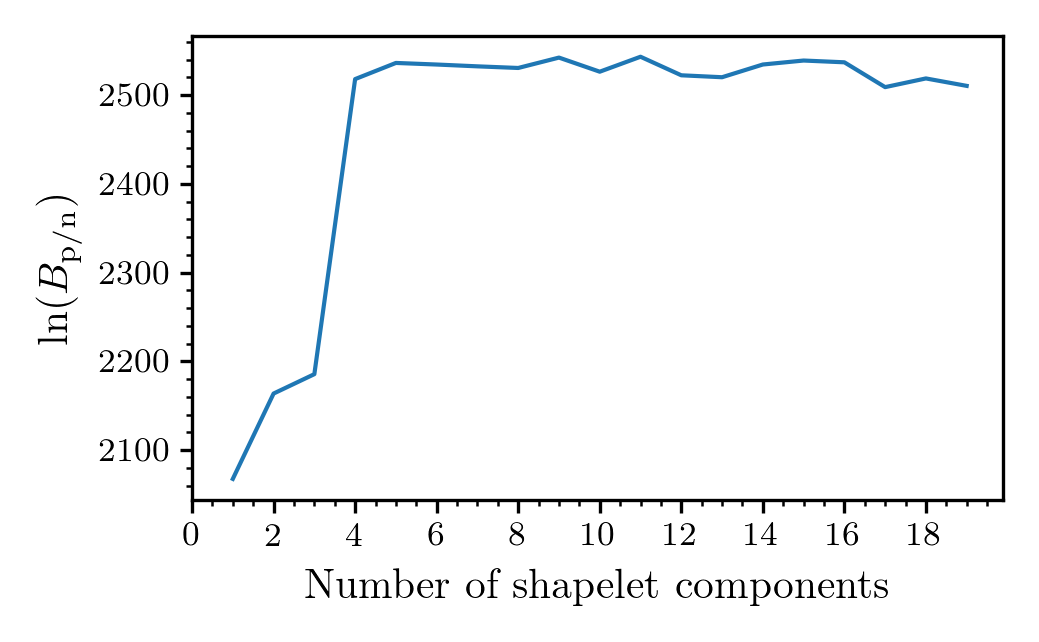}
    \caption{Bayes factor for the pulse vs null applied to simulated data similar to the Vela pulsar.}
    \label{fig:bayes_factor}
\end{figure}

\begin{figure}
    \centering
    \includegraphics[width=0.45\textwidth]{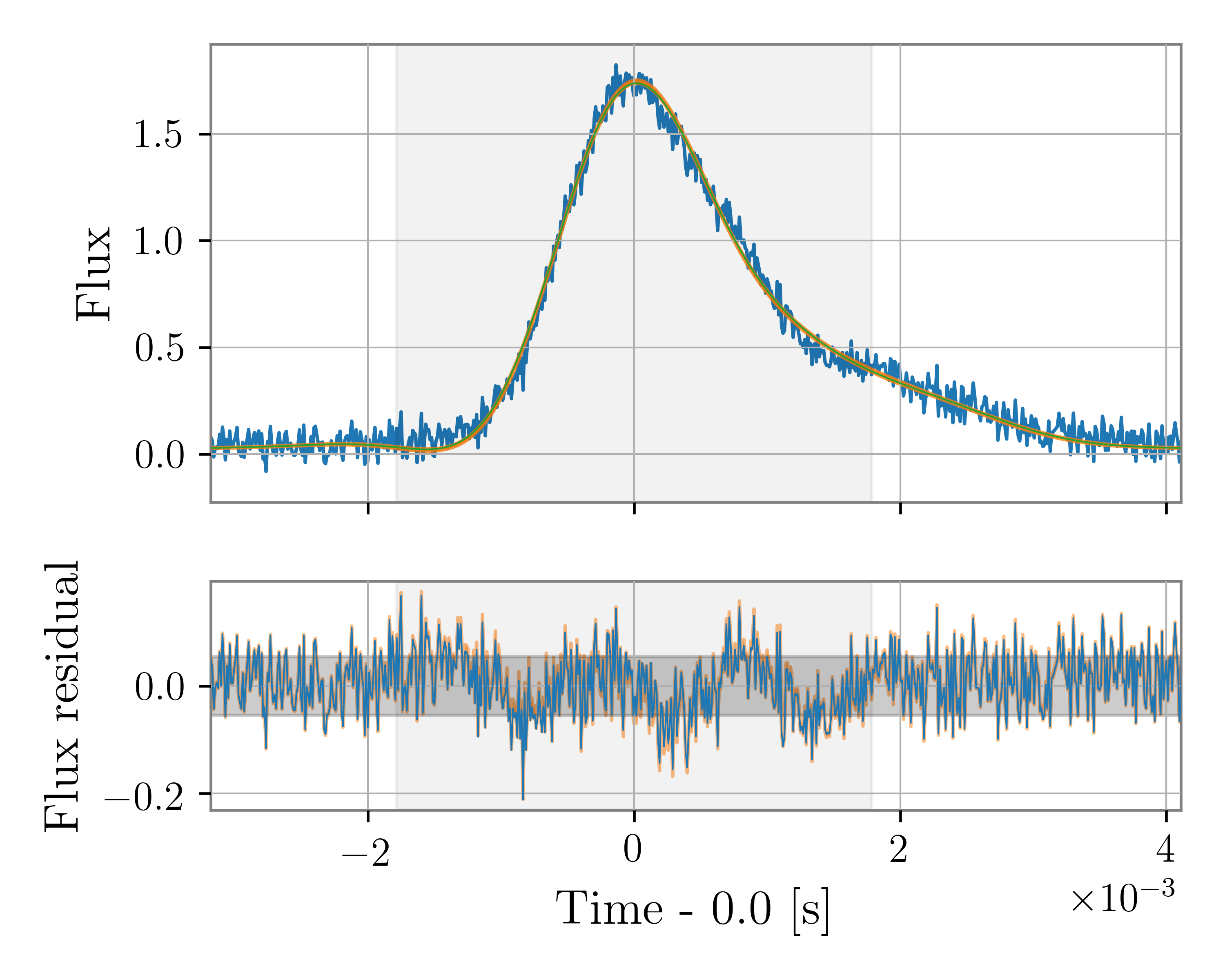}
    \caption{Top panel: the simulated data (blue), maximum likelihood fit (green), and 90\% confidence interval (C.I.) of the fit (orange) for the 20-component shapelet model fit. Bottom panel: the residual after removing the maximum likelihood (blue) and 90\% C.I uncertainty (orange). A grey region indicates the user-settable prior region for the pulse time of arrival.}
    \label{fig:fit}
\end{figure}

\begin{figure}
    \centering
    \includegraphics[width=0.45\textwidth]{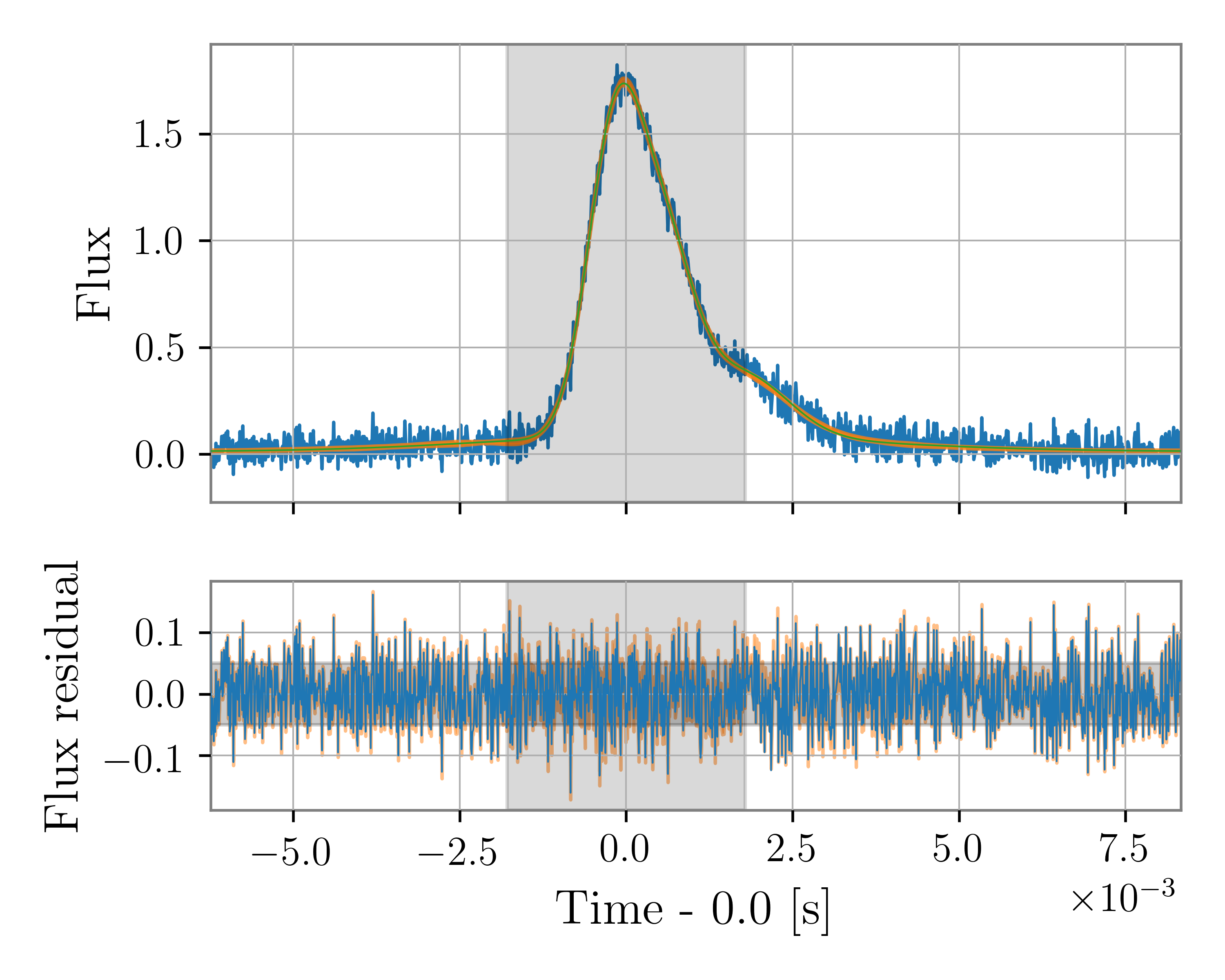}
    \caption{Top panel: the simulated data (blue), maximum likelihood fit (green), and 90\% confidence interval (C.I.) of the fit (orange) for the 3-shapelet model, each with 5 components. Bottom panel: the residual after removing the maximum likelihood (blue) and 90\% C.I uncertainty (orange). A grey region indicates the user-settable prior region for the pulse time of arrival.}
    \label{fig:fit-fixed}
\end{figure}
To demonstrate the behaviour of \kookaburra with a varying number of components,
we create a simulated data set (shown later in \cref{fig:fit-fixed}) based
on the profile of the Vela pulsar. We fit this data with a model consisting of a single
shapelet flux without any base flux (the simulated data does not include a base
flux). In \cref{fig:bayes_factor}, we plot $B_{\rm p/n}$, the pulse-to-null
Bayes factor as a function of the number of components of the shapelet flux.
In this example, below six components, vast improvements in the fit are
achieved by the addition of extra components. Above six, modest improvements
are made as more subtle features get fit, but overall the behaviour plateaus.
Eventually, the Bayes factor will turn over and start to decrease as additional
components fail to improve the fit, but incur extra losses from the increased
prior space (this is the aforementioned Occam penalty). 

In \cref{fig:fit}, we show the output of \kookaburra, for the 20-component
model fit to the simulated data. From this fit, it is clear that we have not
found a perfect fit for the data. In particular, the residual demonstrates
structure indicating the existence of an improved model fit. It is possible
that increasing the number of components will eventually resolve this
under-fitting. However, we find in this case that a model with 3 shapelets,
each having 5 components improves the fit; see \cref{fig:fit-fixed}. Here,
we simply demonstrate that a reasonable fit can be achieved, studying the
optimal choice of model is a future research project.

\textbf{Sensitivity:} To build intuition about the behaviour of the Bayes factor, we simulate a Vela-like pulsation (similar to the simulated data in \cref{fig:fit}), but we vary the peak-amplitude of the pulsation relative $\sigma$, the standard deviation of the simulated background noise.

In \cref{fig:sensitivity}, we plot the recovered Bayes factor for three different search setups (or equivalently, three different priors on the shapelet parameters). The blue curve is the Bayes factor for a search only over the TOA, the shapelet parameters are fixed at the simulation values. As such, the blue curve is similar to a traditional TOA search which uses a fixed-pulse-template. In this case, the median Bayes factor never favours the null model. This is because the Occam factor is small; in this search, we are providing a significant amount of information to the ``pulse'' model about the shape of the expected pulse. This extra information means that for weak pulses, such a search is more sensitive in identifying pulsations. This gain in sensitivity is only robust if the extra information is accurate.
The red curve simulates a case where one knows the ``optimal'' number of shapelet parameters needed, but still must search over the parameters. This is less sensitive than than blue curve, because there is less prior information provided about the shape of the pulse.
Roughly, this search can identify pulses when the peak-amplitude is $\sim 0.4\sigma$. The green curve fits an eight-component shapelet model (to data with only four components). This simulates realistic searches, where we do not typically know the shape (or dimensionality) of the shapelet before fitting. This is less sensitive than the red curve (again the prior-volume is larger) resulting in a minimum detectable peak amplitude $\sim0.6 \sigma$.

Fig.~\ref{fig:sensitivity} demonstrates that, without knowing \emph{a priori} the shape of the pulse, there is always a minimum detectable pulsation (which depends in general on the amount of information provided by the search). Future development could include informative priors in which the shapelet parameters are given prior distributions based on fits to other pulses from the same pulsar.

\begin{figure}
    \centering
    \includegraphics{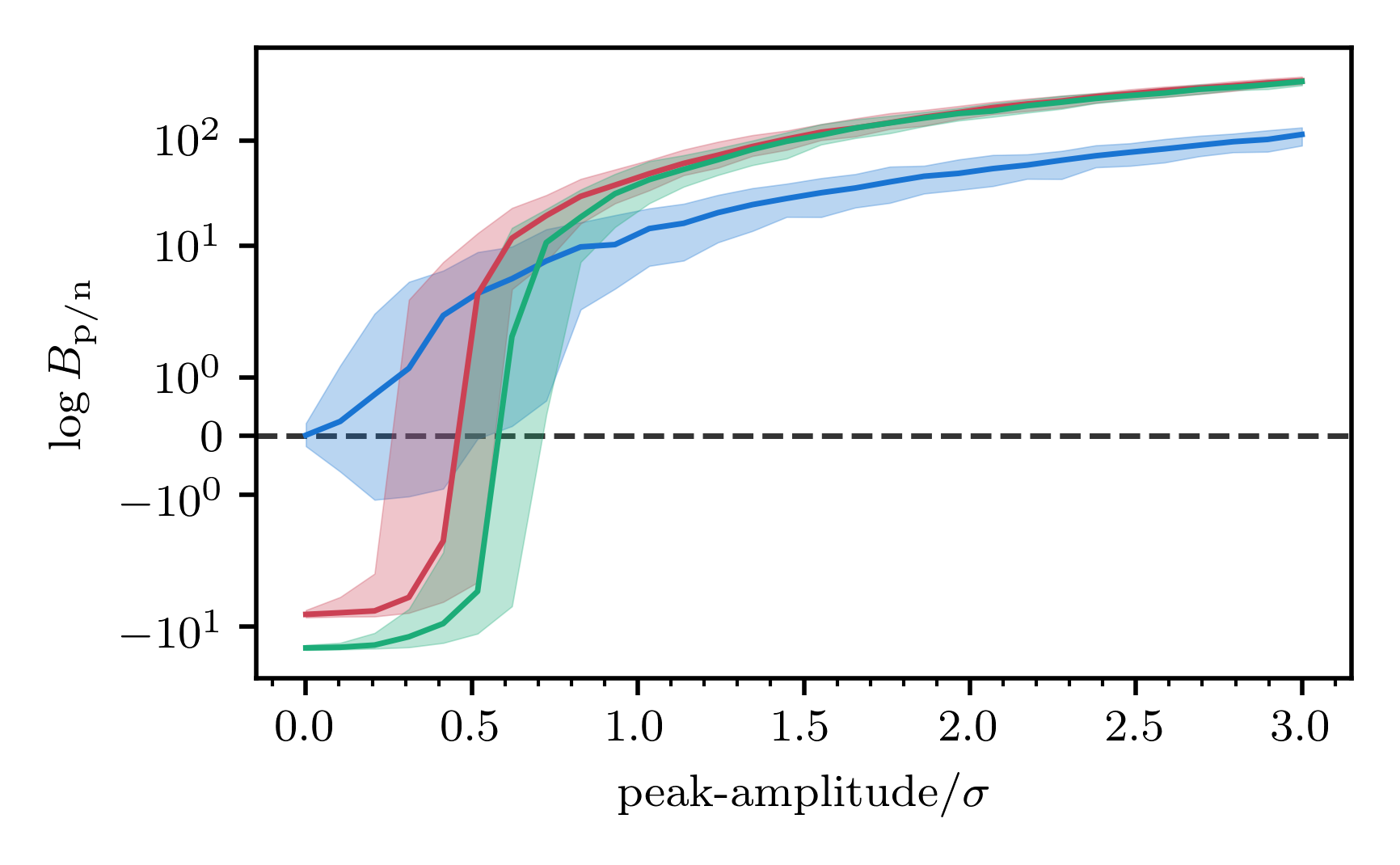}
    \caption{The Bayes factor for three searches applied to a simulated Vela-like pulsation modelled by a four-component shapelet model. We vary the peak-amplitude relative to $\sigma$, the standard deviation of the simulated background noise. In changing the peak amplitude, we keep the shape fixed.
    Blue: a search in which the four-component shapelet pulse is fixed (i.e. the search shapelet matched the simulated shapelet) and only the TOA is estimated. Red: a search in which all four components of the shapelet pulse and the TOA are estimated. Green: a search in which we use an eight-component shapelet and fit all components and the TOA. Solid curves give the median while shaded regions the 90\% uncertainty on the estimated Bayes factors.}
    \label{fig:sensitivity}
\end{figure}

\textbf{Validation}
To validate that the \kookaburra implementation is unbiased, we simulate 100
pulses with a 10-component shapelet and polynomial base-flux of degree 2. We
run \kookaburra on each simulated data set then, in \cref{fig:pp}, we plot a
parameter-parameter (PP) plot \cite{cook:2006} using the posterior samples
calculated for these 100 simulated data sets. The plot demonstrates that each
of the model parameters fit produces a diagonal line on the PP-plot, i.e. the
X\% confidence interval (C.I.) contains the true simulated value X\% of the
time (to within the statistical uncertainties). This demonstrates that for the
default settings (i.e. the \texttt{pymultinest} sampler \cite{buchner:2014} with 1000 live
points), \kookaburra is unbiased in its estimation of the shapelet parameters.
Since the model complexity is user-settable, we cannot guarantee that the
default settings will apply in general and recommend users carefully check for
convergence and, if necessary, run a PP test to validate performance in the
given circumstances of interest.

\begin{figure}
    \centering
    \includegraphics{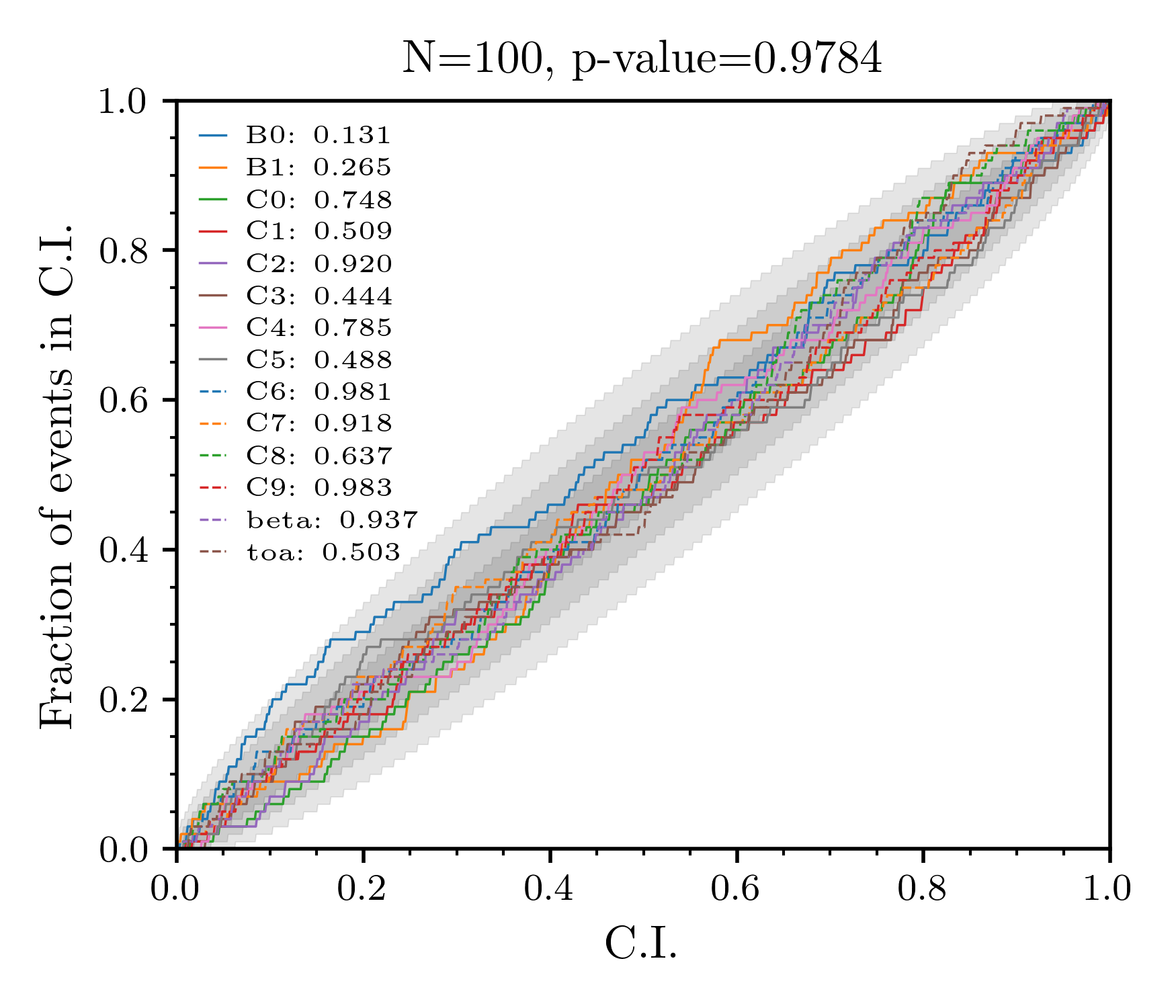}
    \caption{Parameter-parameter test for simulated data with 10-component shapelet and base flux of degree 2.}
    \label{fig:pp}
\end{figure}

\end{methods}

\end{document}